\documentstyle[preprint,aps]{revtex}
\tighten
\draft
\begin{document}
\title{Estimate of Minimal Noise in a Quantum Conductor}
\author{Hyunwoo Lee, L.S.Levitov\cite{*}}
\address{Massachusetts Institute of Technology,\\
12-112, Department of Physics,\\
77 Massachusetts Ave., Cambridge, MA 02139}
\maketitle

\begin{abstract}
We study zero temperature fluctuations of charge  flow
in  a  metallic loop induced by time dependent magnetic flux, and solve
for the optimal way of varying flux in order to minimize  noise.
Optimal time dependence of the flux is a sum of ``solitons,'' each
corresponding  to quantized flux change. Minimal noise coincides
with that for the  binomial  distribution  with  the  number  of
attempts   equal   to  the  number  of  solitons  and  with  the
probabilies defined by the scattering matrix of the system.

\end{abstract}
\narrowtext

In the theory of
noise in quantum conductors it is usually regarded as  a
characteristic  of
transport complementary to conductance. During last years the
literature   on   quantum   noise  was  mostly  concentrated  on
expressing it through transmission  coefficients  of  conduction
channel\cite{1,2,3}.  Having the noise  written  as  a sum of
contributions from different channels allows one  to  relate  it
with the conductance  for  which such representation is known for a
while, and also to compare noise in the two limits, quantum  and
classical.  Comparison with classical shot noise leads
to the understanding of quantum reduction of the noise, an important
effect of Fermi statistics\cite{4,5,6}.

Beyond this there are several other interesting properties of quantum
noise.   One   is the issue of probability  distribution  of  charge
fluctuations.  The  question arises   naturally  from   the
comparison  with the Poisson statistics of classical shot noise.
In the quantum case the statistics was  found  to  be  binomial
with   probabilities   of  outcomes  related  with  transmission
coefficients of elastic scattering in the system, and  with  the
number of attempts weakly fluctuating near $eVt/h$, where $V$ is
voltage   applied   to  the  system  and  $t$  is  the  time  of
measurement$^{7}$.

Another curious property of quantum noise described recently  is
its   phase   sensitivity  named  `non-stationary  Aharonov-Bohm
effect'$^{8}$. As its simplest realization  one  can  consider
metallic  wire  bent  in  a  loop  with magnetic flux applied to
it$^{9}$. In this setup, when the flux  is  changed  from  one
constant value to another during fixed time interval $\tau$, the
fluctuations  of  charge  measured  during  much longer time $t$
contain a term  proportional  to  $\sin^2(\pi\Delta\Phi/\Phi_0)\ln
t/\tau$,   where  $\Delta\Phi$  is  total  change  of  flux  and
$\Phi_0=hc/e$. Both properties of  this  term,  the  logarithmic
divergence  and  the  periodicity  in $\Delta\Phi$ can be put in
connection with the orthogonality catastrophe theory by relating
$\Phi(t)$ with the phase of forward scattering of the part of the
wire threaded by the flux. In a different setup, involving  same
loop  geometry  but the flux $\Phi(t)$ varying periodically with
time, the phase sensitivity leads to singularities of the  noise
at  integer  $eV/\hbar\Omega$,  where  $V$  is  dc  voltage  and
$\Omega$  is the frequency of the ac flux$^{8}$. The strengths
of the singularities  are  oscillating  functions  of  the  flux
amplitude and are independent of the frequency $\Omega$.

The question we address in this paper is about  optimal  way  of
changing flux that minimizes induced noise.  It  is  clear  from
what  has  been said that for achieving minimum of the noise one
should change the flux by integer amount,
\begin{equation}
\Delta\Phi=\Phi_{t=\infty}-\Phi_{t=-\infty}=\ n\Phi_0\ ,
\end{equation}
in order to suppress the logarithmic term. However, since for  a
given  $\Delta\Phi$  the  noise  depends  on the actual function
$\Phi(t)$,  not  just  on  $\Delta\Phi$,  we  have a variational
problem to solve for the noise  as  a  functional  of  the  time
dependence of the flux. This functional was derived for a single
channel ideal conductor with one  localized  scatterer$^9$.  For
$T=0$ it is given by
\begin{equation}\langle\!\langle Q^2\rangle\!\rangle=\ {{\rm g} e^2
\over2\pi} D(1-D) \int\ {\Big |}
\int e^{i\phi(t)+i\omega t} dt {\Big |}^2
|\omega| {d\omega\over 2\pi} \ ,
\end{equation}
   where $\phi(t)=2\pi\Phi(t)/\Phi_0$, $D$ is transmission
coefficient, and ${\rm g}$ is spin degeneracy. We shall
study the variational problem (2) with the boundary condition
(1) and show that its general
solution has the form of a sum of `solitons':
\begin{equation}
\Phi(t)=\pm{\Phi_0\over \pi} \sum_{k=1}^n \tan^{-1} {\big (}
{t-t_k\over\tau_k} {\big )} \ , \ \tau_k>0,
\end{equation}
  where $t_k$ and $\tau_k$ are arbitrary constants. Under
condition (1) any time dependence of the form (3) gives absolute
minimum to the noise:
\begin{equation}
{\rm min}[\ \langle\!\langle Q^2\rangle\!\rangle\ ] =\
{\rm g} e^2 D(1-D)\ |n| \ ,
\end{equation}
  For an optimal time dependence of the  voltage
$V=-\partial\Phi/c\partial t$
therefore one  has  a  sum  of  Lorentzian peaks:
\[V(t)=\ \mp{\Phi_0\over c\pi}\
\sum_{k=1}^n{\tau_k\over(t-t_k)^2+\tau_k^2}\ .\]
  In order to compare quantum noise  with  conductance,  let  us
mention  that  average  transmitted  charge  $\langle\!\langle Q
\rangle\!\rangle$ equals ${\rm  g}  e  D\Delta\Phi/\Phi_0=  {\rm
g}{e^2\over h}D\int V(t)dt$, in accordance with Ohm's law, i.e.,
there  is  no  particular  dependence on the way the flux change
$\Delta\Phi$ is realized.

The result (3),(4) has a simple interpretation in terms of the
binomial statistics picture of charge fluctuations. For the
binomial distribution with probabilities of outcomes $p$ and
$q$, $p+q=1$, and with the number of attempts $N$, second moment
is known to be equal to $pqN$. The comparison with Expr.(4) suggests to
attribute to $n=\Delta\Phi/\Phi_0$ the meaning of the number of
attempts. This interpretation is supported by the structure of
the function (3) consisting of $n$ terms, each corresponding to
unit change of flux. A remarkable property of the function (3)
is its separability, expressed clearly both in the form of the
terms and in the way the parameters $t_k,\tau_k$ enter the
expression. Let us note that by making some of the $t_k$'s
close to each other
one can have an overlap in time of the `attempts'. The overlap,
however, does not change the fluctuations (4). The situation
reminds the one with solitons in integrable non-linear systems,
or with non-interacting instantons in integrable field theories.
Also, the absence of interference is interesting in the context
of coherent nature of transport in this system: after all, we
are just talking about scattering off of a time-dependent
potential. Perhaps, proper interpretation of this effect should
be sought in establishing relation with the theory of
coherent states that have the property to eliminate to some
extent the quantum mechanical interference.

Let us now turn to the variational problem. It is convenient to
do the integral over $\omega$ first and to rewrite (2) as
\begin{equation}
\langle\!\langle Q^2\rangle\!\rangle=\ -{A\over\pi}\ \int\int
{e^{i\phi(t)-i\phi(t')}\over(t-t')^2}dtdt'\ ,
\end{equation}
  where $A= {{\rm g}e^2\over2\pi}D(1-D)$. In order to avoid
divergence at $t=t'$ the denominator in (5) should be understood
as
\begin{equation}
{1\over 2}{\Big [}{1\over(t-t'+i\delta)^2}+
{1\over(t-t'-i\delta)^2}{\Big ]}\ ,
\end{equation}
  the condition that one obtains by introducing
regularization in (2): $|\omega| \rightarrow |\omega|
e^{-|\omega|\delta}$. By considering variation of the
functional (5) we have the equation for an extremum:
\begin{equation}
{\rm Im} {\Big [} e^{i\phi(t)} \int {e^{-i\phi(t')}
\over(t-t')^2}dt {\Big ]}=\ 0\ .
\end{equation}
   By using Cauchy formula one checks that the functions
\begin{equation}
e^{i\phi(t)}=\prod_{k=1}^n{t-\lambda_k\over t-{\bar\lambda}_k}
\ ,\ \lambda_k=t_k+i\tau_k\ ,
\end{equation}
  satisfy (7) provided all $\tau_k$ are of same sign. Obviously,
the functions (8) are just another form of (3).

It remains to be shown that the functional reaches its minimum
on the solutions (8). To prove it we proceed in the following
steps. Let us write $e^{i\phi(t)}$ as
\begin{equation}
e^{i\phi(t)}=f_+(t)+f_-(t)\ ,
\end{equation}
    where $f_+(t)$ and $f_-(t)$ are bounded  analytic  functions
of  complex $t$ in the upper and lower half plane, respectively.
Representation (9) exists  for  any  non-singular  function  and
defines  the functions $f_+$ and $f_-$ up to a constant. Then we
substitute Expr.(9) in (5) and  apply  Cauchy  formula  for  the
derivative,
\[\dot f_\pm(t)=\pm{i\over2\pi}\oint{f_\pm(t')dt'\over(t-t')^2}\ , \]
where  the  contour of integration is chosen in the halfplane of
analyticity of $f_+$ or $f_-$, respectively. Thus one gets
\begin{equation}
\langle\!\langle Q^2\rangle\!\rangle=\ - i A\ \int
({\bar f}_+{\dot f}_+ - {\bar f}_-{\dot f}_-) dt\ .
\end{equation}
   On the other hand,
\begin{equation}
n\ =\ {1\over 2\pi i} \int e^{-i\phi(t)} {d \over dt}
e^{i\phi(t)}dt= \ -{i\over 2\pi}\int ({\bar f}_+{\dot f}_+ +
{\bar f}_-{\dot f}_-) dt\ ,
\end{equation}
   where the last equality is a result of substituting (9) and using
$\int \bar f_+\dot f_- =\int \bar f_-\dot f_+=0$ that follows from
Cauchy theorem. Now, Expr.(10) can be rewritten through Fourier
components of $f_+$ and $f_-$ as
\[\langle\!\langle Q^2\rangle\!\rangle=\ A\
\int\limits_0^\infty (|f_+(\omega)|^2+|f_-(-\omega)|^2)
\omega{d\omega\over2\pi} \ ,\]
  thus demonstrating positivity of both terms in  (10).  (It  is
used  that  $f_+(\omega)=f_-(-\omega)=0$  for  $\omega<0$.) With
this, by comparing (10) and (11) we obtain
\begin{equation}
\langle\!\langle Q^2\rangle\!\rangle\ \ge\ 2\pi A\ |n|\ .
\end{equation}
Equality in (12) is reached only when either $f_+(t)$ or $f_-(t)$
vanishes. Therefore, to obtain the minimum one has to  take  the
functions  $e^{i\phi(t)}$  that  are  regular in one of the half
planes. This remark is sufficient to see that the functions  (8)
form a complete family of solutions.

It is worth mentioning that the method used to derive (12) copies
almost entirely  the procedure of  derivation  of  the  `duality'
condition in the theory of instantons. Like in other  situations
where   the  duality  condition  holds  our  `solitons'  do  not
interact:  $\langle\!\langle   Q^2\rangle\!\rangle$   shows   no
dependence  on  the  parameters $\lambda_k$ of the solution (8).
Among numerous field theories that allow for exact  solution  of
the  instanton  problem  the  one  most  similar to our case is the
theory of classical Heisenberg ferromagnet in  $D=2$.  For  this
case  the  instantons  were  found by mapping the order
parameter space (i.e., the unit  sphere)
on  the  complex  plane$^{10}$.
Duality  condition  was  shown  to take the form of the  constraint of
analyticity or anti-analyticity of the  mapped  order  parameter
function (compare with the above derived condition $f_+=0$ or $f_-=0$).
Multi-instanton  solutions  were  given  as  products  of single
instanton  solutions  (cf.  Expr.(8)).  This  analogy  obviously
deserves more attention.

At this point let us examin an interesting non-optimal time
dependence of the flux, the sum of two solitons with opposite
charge:
\begin{equation}
\Phi(t)={\Phi_0\over \pi} {\big [} \tan^{-1} {\big (}
{t-t_1\over\tau_1} {\big )} - \tan^{-1} {\big (}
{t-t_2\over\tau_2} {\big )} {\big ]} \ ,
\end{equation}
  $\tau_{1,2}>0$. This function corresponds to $e^{i\phi(t)}$ of
the form (8) but with the poles in both half planes. In this
case $\Delta\Phi=0$ and thus $\langle\!\langle
Q\rangle\!\rangle=0$, so ${\rm min}[\ \langle\!\langle
Q^2\rangle\!\rangle\ ] =0$. With the function (13), however, one
finds
\begin{equation}
\langle\!\langle Q^2\rangle\!\rangle=\ 4\pi A\ {\big |}
{\lambda_1-\lambda_2\over \lambda_1-\bar\lambda_2} {\big |}^2\ ,
\end{equation}
  where   $\lambda_{1,2}=t_{1,2}+i\tau_{1,2}$.   For   different
values  of  the  parameters  $t_{1,2}$,  $\tau_{1,2}$  Expr.(14)
interpolates   between   two   trivial   limiting   cases:   (i)
$\langle\!\langle  Q^2\rangle\!\rangle\to0$  when  the  two flux
steps in (13) have almost equal  duration  and  almost  overlap;
(ii)  $\langle\!\langle Q^2\rangle\!\rangle\to 4\pi A$, when the
flux steps either differ strongly in their duration  or  do  not
overlap.  In the case (ii) the noise is two times bigger
than the noise due to a single step, as it should be.

We see that when $\Delta\Phi/\Phi_0$ is of the order  of  one  a
non-optimal  time  dependence $\Phi(t)$ can considerably enhance
the  noise.  It is not the case, however, for $\Delta\Phi/\Phi_0
\gg 1$. This limit was analyzed in our previous paper and it was
found that when $\Phi(t)$ is a monotonous function the result
\begin{equation}
\langle\!\langle Q^2\rangle\!\rangle=\ {\rm g}e^2D(1-D)
|\Delta\Phi/\Phi_0|\
\end{equation}
   is  quite  accurate$^{9}$.  By  studying  several examples we
concluded that relative correction is small. However,  there  is
more  to  say about how big the correction can be. Let us derive
the result (15) by another method that allows to trace out the order
of magnitude of higher order terms. For that  let  us  take  the
flux  in  the form $\Phi(t)={\Delta\Phi\over2\pi}\phi(t)$, where
$\phi(t)$ is a smooth  monotonous  function,  $\phi(-\infty)=0$,
$\phi(\infty)=2\pi$.  For integer $N=\Delta\Phi/\Phi_0\gg 1$ the
Fourier component of $e^{iN\phi(t)}$ entering  Expr.(2)  in  the
stationary phase approximation is given by
\begin{equation}
\int\limits_{-\infty}^\infty e^{iN\phi(t)+i\omega t} dt =
\sum_k\ \sqrt{2\pi i\over N\ddot\phi(t_k)} e^{iN\phi(t_k)+i\omega
t_k}\ +\ {\rm O}(N^{-3/2})\ ,
\end{equation}
  where   $t_k$'s   are   real   solutions   of   the   equation
$N\dot\phi(t)+ \omega = 0$. Then we can write
\[{\big |}
\int\limits_{-\infty}^\infty e^{iN\phi(t)+i\omega t} dt{\big |}^2=
\sum_k\ {2\pi \over N\ddot\phi(t_k)} \ +\ {\rm O}(N^{-2}) \ ,\]
and thus obtain
\begin{equation}
\langle\!\langle Q^2\rangle\!\rangle=\ A\
\int\limits_{-\infty}^\infty \sum_k\ { |\omega| d\omega
\over N\ddot\phi(t_k)}\ +\ ... \ ,
\end{equation}
  where  dots  represent  higher order terms. By differentiating
both sides of the equation $N\dot\phi(t)= -\omega$ one finds the
relation  $d\omega=-N\ddot\phi(t_k)dt_k$,   which   means   that
$|\omega| d\omega /\ddot\phi(t_k) = - |\dot\phi(t_k)| dt_k$, and
therefore      the      integral     in     Expr.(17)     equals
$N\int_{-\infty}^\infty   d\phi=2\pi\Delta\Phi/\Phi_0$.    Since
$|\omega|d\omega$  scales  as $N^2$, the correction to Expr.(17)
can be evaluated as ${\rm O}(1)$, i.e., it is of  the  order  of
one  for any $N$. This means that relative accuracy of Expr.(15)
is ${\rm O}(1/N)$.

A more intuitive way to understand the accuracy of Expr.(15) is to note
that for a given $n$ the number of parameters in the optimal
flux dependence (3) is $2n$, i.e., half of them are `redundant'.
Because of that any smooth monotonous function with sufficiently
large variation $\Delta\Phi$ can be rather accurately  approximated  by  a
function  of  the  form (3), and therefore the noise exceeds the
lower  bound  just  slightly.

An  implication  of  this  result  for  the  binomial statistics
picture is as follows. As it was  discussed  above  there  is  a
(conjectured) correspondence of the terms of Expr.(3) and of the
attempts.  The deviation from the binomial distribution, that of
course  should exist for a non-optimal flux function, in the case
of a smooth $\Phi(t)$ will remain bounded  as  $\Delta\Phi$
increases taking integer values. In more precise terms,
accurate distribution
will  be  written  as  a mixture of
binomial distributions with different
numbers $N$ of  attempts,  $P(m)=\sum_N\rho_NP_N(m)$,
where $P_N(m)=p^mq^{N-m}C_N^m$. The estimated correction implies
that the distribution of attempts $\rho_N$ has finite variance
in the limit $N=\Delta\Phi/\Phi_0\to\infty$.

Before we close let us  mention  that  in  order  to  apply  our
results  in  the  case  of a  mesoscopic  metallic  conductor with
disorder, which is described by many  conducting  channels,  one
just needs to  replace  $D(1-D)$  by  $\sum_nT_n(1-T_n)$,  since
different   scattering   channels   contribute   to   the  noise
independently. The condition of validity of our  treatment  then
is  that the variation of the flux is sufficiently slow, so that
${\rm min} [\tau_k ] \gg \hbar/E_c$, the time of diffusion  across  the
sample. However, at non-zero temperature one also has to satisfy
the  condition  $\tau_k\ll\hbar/T$,  the time of phase breaking.
So, the temperature interval  where our estimate of the noise holds is
$T\le E_c$.

In conclusion, we studied dependence of the noise in a quantum
conductor on the shape of voltage pulse applied to it and found
optimal time dependence that provides minimum of the noise for
given average transmitted charge. Solution displays interesting
analogy with the problem of instantons in the field theories
obeying `duality' condition. Optimal time dependence is a sum of
Lorentzian peaks of voltage, each corresponing to a `soliton' of
flux. The change of flux for a soliton is equal to the flux
quantum $\Phi_0$. The solitons are interpreted in terms of the
binomial statistics picture of charge fluctuations as attempts
to transmit electrons, one electron per each soliton.

\acknowledgements
We are grateful to Gordey Lesovik for the influence he had on
this work. Research of L.L. is partly supported by Alfred Sloan
fellowship.

\end{document}